\documentclass[prb,twocolumn,aps,floats,floatfix,superscriptaddress]{revtex4}
\usepackage{epsfig}
\usepackage{colordvi}
\usepackage{graphicx}
\usepackage{amssymb}
\usepackage{color}

\begin{document}
%\draft
\title{Electron emission from plasmonically induced Floquet bands at metal surfaces}
\author{Branko Gumhalter\footnote{Corresponding author. Email: branko@ifs.hr}}
\affiliation{Institute of Physics, HR 10000 Zagreb, Croatia}
\author{Dino Novko}
\affiliation{Institute of Physics, HR 10000 Zagreb, Croatia}
\author{Hrvoje Petek}
\affiliation{Department of Physics and Astronomy, University of Pittsburgh, Pittsburgh, Pennsylvania 15260, USA}

\begin{abstract}
 We explore the possibility of existence of plasmonically generated electronic Floquet bands at metal surfaces by studying the gauge transformed  electron-surface plasmon interaction in the prepumped plasmonic coherent state environment. These bands may promote non-Einsteinian electron emission from metal surfaces exposed to primary interactions with strong electromagnetic fields. Resonant behaviour and scaling of emission yield with the parent electronic structure and plasmonic state parameters are estimated for Ag(111) surface. Relative yield intensities from non-Einsteinian emission channels in photoelectron spectra offer the means to calibrate the mediating plasmonic fields and therefrom ensuing surface Floquet bands.
\end{abstract}

\date{\today}
\maketitle

\newcommand{\bq}{\begin{equation}}
\newcommand{\eq}{\end{equation}}

\newcommand{\barr}{\begin{eqnarray}}
\newcommand{\earr}{\end{eqnarray}}

\renewcommand{\bibnumfmt}[1]{[#1]}
\renewcommand{\citenumfont}[1]{#1}

\def\brho{{\hbox{\boldmath$\rho$}}}
\def\bvarepsilon{{\hbox{\boldmath$\varepsilon$}}}
\def\bepsilon{{\hbox{\boldmath$\epsilon$}}}
\def\bnu{{\hbox{\boldmath$\nu$}}}
\def\bxi{{\hbox{\boldmath$\xi$}}}
\def\bcalH{{\hbox{\boldmath$\cal{H}$}}}
\def\bcalL{{\hbox{\boldmath$\cal{L}$}}}
\def\bcalW{{\hbox{\boldmath$\cal{W}$}}}
\def\bcalP{{\hbox{\boldmath$\cal{P}$}}}
\def\bcalR{{\hbox{\boldmath$\cal{R}$}}}
\def\bpi{{\hbox{\boldmath$\pi$}}}
\def\bcalA{{\hbox{\boldmath$\cal{A}$}}}

\section{Introduction}
\label{sec:introduction}

In a series of publications we have studied the various aspects of nonlinear electronic response of metals to external electromagnetic (EM) fields as revealed by spectroscopic analysis of multiphoton photoemission yields.[\onlinecite{TM,ReutzelPRX,MarcelPRL,AndiNJP,ACSPhotonics,plasPE,AndiPRB22}]. In Ref. [\onlinecite{plasPE}] we have specifically focused on the occurence of electron emission channels mediated by bulk plasmons generated subsequently to the primary interactions of  electron system with the EM field.[\onlinecite{GW+C}] A peculiar characteristics of these channels is the absence of linear scaling of the emitted electron energy $\epsilon_{f}$ with the multiples $n$ of the absorbed photon energy $\hbar\omega_{x}$ that would be in accord with generalized Einstein's relation 
\bq
\epsilon_{f}=n\hbar\omega_{x}+\epsilon_{b},
\label{eq:Einstein}
\eq
that describes direct multiphoton-induced electron excitations from the initial state with binding energy $\epsilon_{b}$.[\onlinecite{plasPE,UebaGumhalter}]
 Such emergence of nonlinear non-Einsteinian yields starting from the bulk plasmon onset energy were detected in two-photon photoemission (2PP) from (111), (100) and (110) surfaces of silver[\onlinecite{plasPE}] and their energetics is summarized in Fig. \ref{AllAg}. Similar observation regarding one-plasmon mediated electron emission was previously reported[\onlinecite{Horn}] albeit differently interpreted.

In the present work we investigate a complementary mechanism of non-Einsteinian electron yield via the surface plasmon-assisted emission channels[\onlinecite{Hommelhoff,Elezzabi,Bevillon,Hartelt}] building on the plasmonically induced surface Floquet bands.[\onlinecite{Floquet,Floquet1,Floquet2,MarcelNComm,NuskePRR2020}] We shall demonstrate the formation of such bands in electron systems subject to pumping of coherent plasmonic states by strong external perturbations.[\onlinecite{plasPE}]   
  To this end we first introduce in Sec. \ref{sec:System} a gauge transformation of the standard model Hamiltonian describing electron interaction with surface plasmons (SP) to express it in the form of electron coupling to SP vector field. In this gauge the electron wavefunction is amenable to a Volkov Ansatz-type of representation which we derive through a sequence of physically motivated approximations. Based on this wavefunction we show that the SP field generates Floquet sidebands upon 
the quasi-two dimensional (Q2D) electron bands on metal surfaces. In Sec. \ref{sec:Floquet} we study electron emission from excited states of surface Floquet bands and derive the appropriate transition rates.  In Sec. \ref{sec:discussion} we apply the obtained results to make semiquantitative estimates of SP-assisted electron emission from Q2D Floquet bands on Ag(111) surfaces and analyze optimal conditions for and limitations of detection of such effects. We also hint at the experimental verification of predicted phenomena as well as on gaining information on plasmonic coherent states implicated in the process. In Conclusion  section we reiterate basic concepts underlying the developed theory of nonlinear plasmonically induced electron emission rates and yields from Q2D surface Floquet bands. We propose that these be used to recognize and calibrate plasmonic distributions excited by strong external fields. Extensions of the developed concepts to systems likely to exhibit similar effects are briefly indicated.

\section{Model description of electrons coupled to surface plasmons}
\label{sec:System} 
\subsection{Gauge-dependent representations of electron-surface plasmon interaction}
\label{sec:gauges}

We start from a quantum description of electron-SP interaction in a metal based on a standard simple model Hamiltonian $H$ comprising the component that describes the unperturbed electron and SP system and the coupling of electron charge to plasmonic polarization field[\onlinecite{PlasGauge}], viz.
\bq
H=H_{0}^{e}+H_{0}^{pl}+V=H_{0}^{syst}+V.
\label{eq:Hsyst}
\eq
$H_{0}^{e}$ describes an electron in a crystal band, $H_{0}^{pl}$ is the Hamiltonian of  unperturbed surface plasmon field, and $V$ describes their interaction. The electron part reads
\bq
H_{0}^{e}=\frac{{\bf p}^2}{2m}+v({\bf r}),
\label{eq:H0el}
\eq
where ${\bf p}=({\bf P},p_z)$ and ${\bf r}=(\brho,z)$ are the electron momentum and radius vector expressed in cylindrical coordinates, respectively, with $z$  measured perpendicular to the surface at $z=0$.  $m$ is the bare electron mass and $v({\bf r})$ is the effective one-electron crystal potential.  ${\bf r}$ and ${\bf p}$ are the conjugate noncommuting operators satisfying $[{\bf r},{\bf p}]=i\hbar$. Employing  second quantization to represent the boson field of surface plasmons characterized by their two-dimensional (2D) wavevector ${\bf Q}$ and dispersion $\omega_{\bf Q}$ we have for the Hamiltonian of unperturbed plasmons 
\bq
H_{0}^{pl}=\sum_{\bf Q} \hbar\omega_{\bf Q}\hat{a}_{\bf Q}^{\dag}\hat{a}_{\bf Q}=\sum_{\bf Q} \hbar\omega_{\bf Q}\hat{n}_{\bf Q}. 
\label{eq:H0pl}
\eq
 Here the plasmon creation and annihilation operators are denoted by $\hat{a}_{\bf Q}^{\dag}$ and $\hat{a}_{\bf Q}$, respectively, and they satisfy the commutation relations  $[\hat{a}_{\bf Q},\hat{a}_{\bf Q'}^{\dag}]=\delta_{\bf Q,Q'}$. The SP number operator is $\hat{n}_{\bf Q}=\hat{a}_{\bf Q}^{\dag}\hat{a}_{\bf Q}$. 

%-----------------  Fig. 1 -----------------------

\begin{figure}[tb]
\rotatebox{0}{\epsfxsize=8.5 cm \epsffile{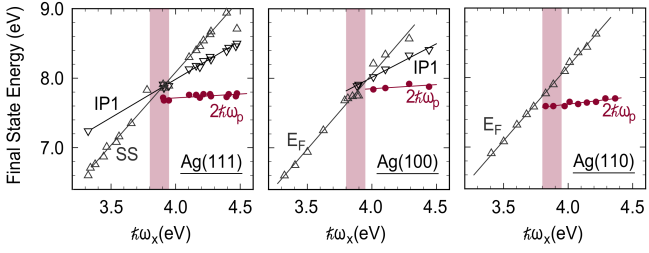}} 
\caption{ Dependence of the measured final state electron energies of 2PP yields on the variation of photon energy $\hbar\omega_x$, as recorded in photoemission from three low index surfaces of Ag [\onlinecite{plasPE}]. The data shown by triangles were measured in the constant inital state mode (CIS) involving the occupied surface state (SS) and the intermediate first image potential state (IP1) on Ag(111), the IP1 and the Fermi level $E_F$ on Ag(100), and $E_F$ on Ag(110). They exhibit standard scaling of 2PP yield energy with $2\hbar\omega_x$ ($n=2$) in emission from the occupied SS and $E_F$, and with $\hbar\omega_x$ ($n=1$) in emission from the intermediate IP1. By contrast, there also appear non-Einsteinian yields that scale with plasmon energy $\sim 2\hbar\omega_p$ above $E_F$ (dots) and  not with the multiples of the radiation field energy $\hbar\omega_x$. Similar processes involving surface plasmons may also be envisaged.}
\label{AllAg}
\end{figure}
%------------------------------------------------------

The electron-SP coupling is described by
\bq
V=\sum_{\bf Q}V_{\bf Q}e^{i{\bf Q}\brho-Q|z|}(\hat{a}_{\bf Q}+\hat{a}_{\bf -Q}^{\dag}),
\label{eq:V}
\eq
with $Q=|{\bf Q}|$, and 
\bq
V_{\bf Q}=\left(\frac{\pi e^{2}\hbar\omega_{\bf Q}}{Q L^2}\right)^{1/2}=\left(\frac{\pi e^{2}\hbar\omega_{\bf Q}}{Q a_{B}^2}\right)^{1/2}\left(\frac{a_{B}}{L}\right)
\label{eq:V_Q}
\eq
where $e$ is the electron charge and $L$ is the SP field quantization length in the $(x,y)$-plane. For later convenience we have here factorized the ratio of the atomic unit of length $a_{B}=\hbar^{2}/me^{2}$ (Bohr radius) and $L$. Expression (\ref{eq:V_Q}) is valid in the long wavelegth limit $Q\rightarrow 0$ in which surface plasmons are well defined stable excitations.
We have left out from the present model the weaker coupled accoustic surface plasmons[\onlinecite{accousticSP}] that may occur in SS-bands, and whose role in multiexcitation processes was addressed in Ref. [\onlinecite{Rivera}]

The historically most frequently used form of the electron-SP interaction (\ref{eq:V_Q}) corresponds to the so-called "length" or "dipole gauge". However, in nonperturbative treatments of the effects of boson fields on electron motion the choice of a different, "velocity" or "radiation gauge", turns out to be more advantageous as it enables pursuing analytic or closed form nonlinear solutions quite far in the descriptions of electron dynamics. The passage to this gauge is achieved by performing on (\ref{eq:Hsyst}) a canonical (unitary) transformation  defined by a nested commutator expansion
\bq
H'=\exp(iS)H\exp(-iS)=H+[iS,H]+\frac{1}{2}[iS,[iS,H]]+ \cdots
\label{eq:HS}
\eq
where the transformation generator $S$ is a hermitian operator depending only on the electron radius vector ${\bf r}$ and the plasmon field momenta $\propto(\hat{a}_{\bf Q}-\hat{a}_{\bf -Q}^{\dag})$, but not on the electron momentum ${\bf p}$, viz.
\bq
S=-i\sum_{\bf Q}\frac{V_{\bf Q}}{\hbar\omega_{\bf Q}}e^{i{\bf Q}\brho-Q|z|}(\hat{a}_{\bf Q}-\hat{a}_{\bf -Q}^{\dag}).
\label{eq:S}
\eq
This gives for the transformed Hamiltonian[\onlinecite{PlasGauge}]
\bq
H'=\frac{({\bf p-A(r)})^2}{2m} +H_{0}^{pl}+v({\bf r})+\Phi({\bf r}).
\label{eq:H'}
\eq
Here ${\bf A(r)}$ with the dimension of momentum is a plasmonic field vector potential acting on the electron at ${\bf r}$
\bq
{\bf A(r)}=\hbar\nabla S=\sum_{\bf Q}\frac{V_{\bf Q}}{\omega_{\bf Q}}({\bf Q},iQ)e^{i{\bf Q}\brho-Q|z|}(\hat{a}_{\bf Q}-\hat{a}_{\bf -Q}^{\dag}), 
\label{eq:A}
\eq
where $({\bf Q},iQ)$ is a vector with lateral and perpendicular to the surface components ${\bf Q}$ and $iQ$, respectively. Hence, the dynamical component of electron-plasmon interaction now acquires the appearance of coupling in the velocity gauge
\bq
V'=-\frac{({\bf p\cdot A(r)+A(r)\cdot p})}{2m}+\frac{{\bf A(r)}^2}{2m}.
\label{eq:V_A}
\eq
The remaining SP-electron interaction
\bq
\Phi({\bf r})=[iS,V]=-\sum_{\bf Q}\frac{V_{\bf Q}^{2}}{\hbar\omega_{\bf Q}}e^{-2Q|z|},
\label{eq:Phi}
\eq
is an instantaneous scalar potential depending only on the electron coordinates and not on plasmon operators. With the coupling (\ref{eq:V}) this is the standard electron image potential. Note that $\Phi({\bf r})$ arises from virtual excitation (creation and subsequent annihilation) of plasmons and therefore represents a shift of the plasmonic ground state energy induced by the electron. Therefore, the effect of  canonical transformation (\ref{eq:HS}) is to eliminate from the new gauge the  interaction (\ref{eq:V}) and replace the operator ${\bf p}$ and the potential $v({\bf r})$ by
\barr
{\bf p}&\rightarrow& {\bf p-A(r)}=\bpi,
\label{eq:p'}\\
 v({\bf r})&\rightarrow& v({\bf r})+\Phi({\bf r})=V_{scal}({\bf r}).
\label{eq:Vscal}
\earr
Thereby the electron-plasmon interaction enters the transformed $H'$ through the vector potential ${\bf A(r)}$ which satisfies the Coulomb gauge
\bq
{\bf \nabla\cdot A(r)}=0.
\label{eq:nablaA}
\eq
This implies that the operators ${\bf p}=-i\hbar\nabla$ and ${\bf A(r)}$ commute when acting on the electron wavefunctions. This property will prove instrumental in their construction.     

\subsection{Plasmonically induced vector potential}
\label{sec:Apotential}

Standard methods for nonperturbative description of electron motion in strong spatially homogeneous external vector fields are: {\it (i)} constructions of the wavefunction based on the Volkov Ansatz[\onlinecite{Volkov}] introduced to describe free electron interactions with strong EM fields[\onlinecite{Truscott1,Truscott2}] and later elaborated in the calculations of transition rates characteristic of photoionization of atomic[\onlinecite{Keldysh1965,Faisal,Reiss1980,Madsen,Faisal2016,Keldysh2017}] and condensed matter  systems[\onlinecite{Keldysh1965,Reiss1977,Yalunin,Kidd}], or {\it (ii)} Fourier analysis of the underlying time dependent Schr\"{o}dinger equation.[\onlinecite{FaisalKaminski97,FaisalKaminski05,Park,PengZhang}] For convenience we resort to the former method elaborated in Refs. [\onlinecite{Keldysh1965,Faisal,Reiss1980,Madsen,Keldysh2017,Reiss1977,Kidd}]. The Volkov-type representation of electronic states in solids naturally leads to the picture of Floquet bands.  

 With the prerequisites from the preceding subsection we can now establish a correspondence between the present model of electron-plasmon coupling embodied in $H'$ (\ref{eq:H'}) and the Volkov-type of  representation of electron states. Assume that a cloud of real surface plasmons has been excited in the system by the action of a strong perturbation so that this gives rise to occupation numbers $n_{\bf Q}$ of  plasmon modes characterized by the wavevector ${\bf Q}$. The occupations  are determined by the plasmon excitation dynamics and various distributions of excited plasmonic states can be constructed once their values or generating functions are known.[\onlinecite{SunkoBG}]
In the currently addressed problem obeying the temporal boundary conditions of photoemission induced by ultrashort pulses the cloud of excited plasmons reaches the form of  a coherent state $|\alpha\rangle=\Pi_{\bf Q'}|\alpha_{\bf Q'}\rangle$.[\onlinecite{plasPE}] Each component $|\alpha_{\bf Q'}\rangle$ satisfies $\hat{a}_{\bf Q'}|\alpha_{\bf Q'}\rangle=\alpha_{\bf Q'}|\alpha_{\bf Q'}\rangle$ where $\alpha_{\bf Q'}$ is the  parameter of coherent state generating function[\onlinecite{Glauber,GlauberPR}]
\bq
|\alpha_{\bf Q'}\rangle=\exp(-|\alpha_{\bf Q'}|^2/2)\sum_{n}\alpha_{\bf Q'}^n/(n!)^{1/2}|n_{\bf Q'}\rangle.  
\label{eq:cohstate}
\eq

Effects of electron dressing by absorption of plasmons constituting the primary pumped coherent state can be studied by replacing ${\bf A(r)}$ in (\ref{eq:H'}) by an equivalent external time dependent vector potential $\bcalA({\bf r},t)$ depending only on the electron coordinates and $\alpha_{\bf Q}$ characteristic of the plasmon distribution. This is constructed by acting with the plasmon absorption component of ${\bf A(r)}$ on the coherent state $|\alpha\rangle$ and complementing it with its hermitian conjugate. This leads to the SP-generated effective vector potential
\barr
\bcalA({\bf r},t)&=&\left[\sum_{\bf Q}\hat{\bf e}_{\bf Q}\frac{Q V_{\bf Q}}{\omega_{\bf Q}}\alpha_{\bf Q}e^{(i{\bf Q}\brho-Q|z|)}e^{-i\omega_{\bf Q}t} + h.c.\right]\nonumber\\
&+&
\left[i\sum_{\bf Q}\hat{\bf e}_{z}\frac{Q V_{\bf Q}}{\omega_{\bf Q}}\alpha_{\bf Q}e^{(i{\bf Q}\brho-Q|z|)}e^{-i\omega_{\bf Q}t} + h.c.\right]\nonumber\\
&=&
\bcalA_{\parallel}({\bf r},t)+\bcalA_{\perp}({\bf r},t).
\label{eq:bcalA}
\earr
Here $\parallel$ and $\perp$ denote the vector components parallel and perpendicular to the surface plane, respectively, and $\hat{\bf e}_{\bf Q}$ and $\hat{\bf e}_{z}$ are the unit  vectors in the direction of ${\bf Q}$ (i.e. $\hat{\bf e}_{\bf Q}={\bf Q}/Q$) and perpendicular to the surface, respectively. For calculational convenience we shall henceforth assume that the surface plasmon excitation mechanism produces indiscriminate occupation of the ${\bf Q'}$-modes constituting the coherent state (\ref{eq:cohstate}). This assumption renders the ${\bf Q}$-summations in $\bcalA({\bf r},t)$ unrestricted by any additional distribution of excited plasmon modes.

 Expression (\ref{eq:bcalA}) can be further simplified by neglecting the effects of weak surface plasmon dispersion which amounts to taking $\omega_{\bf Q}=\omega_{s}$. 
With the same frequency of all plasmon modes propagating on planar surfaces one may also expect their indiscriminate role in forming the coherent states so that $\alpha_{\bf Q}\rightarrow \alpha_{Q}a_{B}/L$ where the factor $a_{B}/L$ appears from the history of pumping of plasmonic excitations via the interaction matrix elements analogous to $V_{\bf Q}$ defined in (\ref{eq:V_Q}). Within the temporal boundary conditions for generating  plasmon coherent states described in Ref. [\onlinecite{plasPE}] (cf. Eqs. (25) and (38) therein) the amplitudes $\alpha_{Q}$ may be taken real. In this limit the components of (\ref{eq:bcalA}) take the forms 
\barr
\bcalA_{\parallel}({\bf r},t)&=&\sum_{\bf Q}\hat{\bf e}_{\bf Q}\alpha_{Q}{\cal A}_{Q}e^{-Q|z|}\cos({\bf Q}\brho-\omega_{s}t),
\label{eq:Aspar}\\
\bcalA_{\perp}({\bf r},t)&=&-\sum_{\bf Q}\hat{\bf e}_{z}\alpha_{Q}{\cal A}_{Q}e^{-Q|z|}\sin({\bf Q}\brho-\omega_{s}t), 
\label{eq:Asperp}
\earr
where  
\bq
\alpha_{Q}{\cal A}_{Q}=\alpha_{Q}\frac{Q V_{Q}}{\omega_{s}}\frac{a_{B}}{L}.
\label{eq:AQ}
\eq
Hence, the in-surface-plane angular dependence in the sums on the RHS of (\ref{eq:Aspar}) and (\ref{eq:Asperp}) manifests solely through the unit vector $\hat{\bf e}_{\bf Q}$ and the argument ${\bf Q}\brho$ of the carrier wave. The presence of two factors $1/L$ in ${\cal A}_{Q}$ (i.e. the one from fraction $a_{B}/L$ and the other from $V_{Q}$) normalize the summands in (\ref{eq:Aspar}) and (\ref{eq:Asperp}) to quantization area. Integrations $\sum_{\bf Q}\rightarrow\frac{L^2}{(2\pi)^2}\int d^2 {\bf Q}$ then render the potentials $L$-independent. 
 
Replacement of ${\bf A}({\bf r},t)$ by the above $\bcalA({\bf r},t)$ in (\ref{eq:H'}) and corresponding removal of $H_{0}^{pl}$ therefrom results in the Hamiltonian describing electron dynamics governed by the static scalar potential (\ref{eq:Vscal}) and the sum of time dependent effective vector fields (\ref{eq:Aspar}) and (\ref{eq:Asperp}). In what follows we shall ignore the dissipative environment because it is not expected to be of significance on the energy scale of Floquet dynamics[\onlinecite{NuskePRR2020}] discussed below.

\subsection{Volkov Ansatz for plasmon-dressed electron wavefunctions at surfaces}
\label{sec:Volkov}

On low index surfaces of metals which exhibit surface projected band gaps the potential $V_{scal}({\bf r})$ defined in (\ref{eq:Vscal}) can support  the set of Q2D surface state (SS) and image potential state (IP) bands. Localization of SS and IP electrons in the direction perpendicular to the surface is only few atomic radii over the image potential well whereas their Q2D  Bloch state dynamics in the lateral direction parallel to the surface is well described in the effective mass approximation.[\onlinecite{FausterSteinmann,Chulkov}] The corresponding one electron wave function describing electron motion in the $s$-th Q2D surface band in the absence of the vector field from $H'$ reads
\bq
\phi_{{\bf K},s}(\brho,z,t)=e^{i{\bf K}\brho}u_{s}(z)e^{-i(\hbar^{2}K^{2}/2m^{*}+E_s)t/\hbar}/\sqrt{L^{2}},
\label{eq:phi_s}
\eq
where $s$ is the surface band index, ${\bf K}$ is an eigenwavevector of the 2D lateral momentum operator ${\bf P}$, $u_{s}(z)$ is the component of electron wavefunction describing its localization in the $s$-state at the surface, and $E_s$ is the electron energy at the $s$-band bottom. The wavefunctions (\ref{eq:phi_s}) satisfy box normalization $\langle \phi_{{\bf K'},s'}|\phi_{{\bf K},s}\rangle=\delta_{\bf K',K}\delta_{s',s}$. The effective electron masses for motion in the lateral and perpendicular to the surface directions in the $s$-th band are denoted by $m^{*}$ and $m_{s}$, respectively. In this context of particular interest are the (111) surfaces of Ag and Cu with well defined SS- and IP-bands, and whose plasmonic response has also been well explored.[\onlinecite{SilkinLazic}] However, since the two-plasmon data from Fig. \ref{AllAg} do not indicate any resonant $\mbox{SS}\rightarrow\mbox{IP}$ transitions we shall exclude from our further considerations the role of IP-states in  plasmonically assisted emission.

The standard Volkov Ansatz for dressed electron wavefunction is based on the assumption of very slow variation of the vector potential across the range of interaction so that for small ${\bf Q}$ the ${\bf Q}\brho$ products determining spatial variation of the corresponding 2D plane waves can be neglected. For isotropic interaction matrix elements $V_{\bf Q}$ this would immediately eliminate (\ref{eq:Aspar}) and one would be left only with (\ref{eq:Asperp}) acting on the electron wavefunction. This passage is demonstrated in Appendix \ref{sec:beyondDA}.

Action of the vector potential on electrons in surface bands renormalizes their dynamics and energetics. $\bcalA({\bf r},t)$ that replaces ${\bf A(r)}$ in $H'$ remains to satisfy (\ref{eq:nablaA}), implying $[{\bf p},\bcalA({\bf r},t)]=0$, and therefore the order of ${\bf p}$ and $\bcalA({\bf r},t)$ is irrelevant in the action of their products on the electron wavefunction. Moreover, since $u_{s}(z)$ are strongly localized at the surface we shall for computational convenience make a replacement 
\bq
\bcalA({\bf r},t)\rightarrow\bcalA(\brho,z_{s},t).
\label{eq:Az_s}
\eq
where $z_{s}$ is the $z$-coordinate of the maximum of SS-state electron density relative to the here relevant dynamical screening plane of Ag(111) surface.[\onlinecite{Liebsch}] This substitution enables the assessment of standard Volkov Ansatz for representation of electron dynamics in Q2D surface bands within the dipole approximation ${\bf Q}\brho\ll 1$ so that all ${\bf Q}\brho$ in the arguments of periodic functions can be neglected relative to the other factors. In this limit $\bcalA_{\parallel}$ vanishes after angular integration over ${\bf Q}$ and we are left only with $\bcalA_{\perp}$ which is $\brho$ independent. This intuitive result is elaborated in  Appendix \ref{sec:beyondDA}. Hence, in this approximation the Volkov Ansatz leading to Floquet states reads (cf. [\onlinecite{Faisal}]) 
\barr
\psi_{{\bf K},s}^{F}(\brho,z,t)
&=&
\exp\left[i\left({\bf K}\brho -\frac{(\hbar{\bf K})^2}{2m^{*}\hbar}t-\frac{E_{s}}{\hbar}t\right)\right]\nonumber\\
&\times&
\exp\left[-\frac{i}{\hbar}\int^{t}dt'\frac{\bcalA_{\perp}^{2}(z_{s},t')}{2m_{s}}\right]\nonumber\\
&\times&
\exp\left[\frac{i}{\hbar m_{s}}\int^{t}dt'\bcalA_{\perp}(z_{s},t')\hat{p}_{\perp}\right]u_{s}(z)
\label{eq:psiVolkovperp}
\earr
with 
\bq
\hat{p}_{\perp}=\hat{p}_{z}=-i\hbar\frac{\partial}{\partial z}
\label{eq:p_z}
\eq
The field-quadratic term in the exponent on the RHS of (\ref{eq:psiVolkovperp}) can be evaluated by exploiting (\ref{eq:Asperp}) and introducing 
\bq
{\cal P}_{\perp}=\sum_{\bf Q}\alpha_{Q}{\cal A}_{Q}e^{-Q|z_{s}|}
\label{eq:calP}
\eq
which has the dimension of momentum, and 
\bq
{\cal E}_{\perp}=\frac{{\cal P}_{\perp}^{2}}{2m_{s}}=2U_{p}
\label{eq:Eperp}
\eq
with the dimension of energy. Using this we obtain
\bq
\exp\left[-\frac{i}{\hbar}\int^{t}dt'\frac{\bcalA_{\perp}^{2}(z_{s},t')}{2m_{s}}\right]\rightarrow \exp\left[-i\frac{{\cal E}_{\perp}}{2\hbar}t+\frac{i{\cal E}_{\perp}}{4\hbar\omega_{s}}\sin(2\omega_{s}t)\right]
\label{eq:A^2}
\eq
in which the first term on the RHS is recognized as the ponderomotive energy shift $U_{p}={\cal E}_{\perp}/2$. Within the present theory it is intrisically positive because it arises from the quadratic coupling term $\bcalA^2({\bf r},t)/2m_{s}$ in the Hamiltonian.  The second term is the double plasmon frequency Floquet term[\onlinecite{ReissPRA42}] weighted by dimensionless two-plasmon absorption amplitude
\bq
\beta_{s}=\frac{{\cal E}_{\perp}}{4\hbar\omega_{s}}=\frac{U_{p}}{2\hbar\omega_{s}}.
\label{eq:beta_s}
\eq
The single plasmon frequency Floquet term[\onlinecite{Keldysh1965,ReissPRA42}] 
arises from the exponent on the RHS of (\ref{eq:psiVolkovperp}) that is linear in $\bcalA_{\perp}$. This term represents a translation operator in the $z$-space[\onlinecite{Faisal,MadsenPRA}] when acting on the wavefunction $u_{s}(z)$, viz.
\barr
& &\exp\left[\frac{1}{m_{s}}\int^{t}dt'{\cal P}_{\perp}\sin(\omega_{s}t)\frac{\partial}{\partial z}\right]u_{s}(z)\nonumber\\
&=&
\exp\left[-Z_{s}\cos(\omega_{s}t)\frac{\partial}{\partial z}\right]u_{s}(z)=u_{s}(z-Z_{s}(t)).
\label{eq:uZs}
\earr
Here 
\bq
Z_{s}={\cal P}_{\perp}/m_{s}\omega_{s},
\label{eq:Z_s}
\eq
and $Z_{s}(t)=Z_{s}\cos(\omega_{s}t)$ with the mean value $\overline{Z_{s}(t)}=0$. Periodic $Z_{s}(t)$ causes jittering of the  $s$-state electron density with the frequency $\omega_{s}$ and phase lag $\pi/2$ relative to the driving plasmon field $\bcalA_{\perp}({\bf r},t)$. Therefore, in the  considered limit the Volkov Ansatz renormalization of the electron wave function is obtained by combining the application of (\ref{eq:A^2}) and (\ref{eq:uZs}) on $u_{s}(z)$, viz. through
\bq
e^{-iU_{p}t/\hbar}\exp\left[i Z_{s}\cos(\omega_{s}t)\frac{i\partial}{\partial z}+i\beta_{s}\sin(2\omega_{s}t)\right]u_{s}(z).
\label{eq:VolkovExp}
\eq
Here the second exponential can be written as  
\barr
& &\exp\left[i Z_{s}\sin(\pi/2-\omega_{s}t)\frac{i\partial}{\partial z}+i\beta_{s}\sin(2(\pi/2-\omega_{s}t))\right]\nonumber\\
&=&
\sum_{n=-\infty}^{\infty} e^{-i n\omega_{s}t}(i)^{n}J_{n}\left(i Z_{s}\frac{\partial}{\partial z},\beta_{s}\right)
\label{eq:generating}
\earr
where we have made use of the generating function for generalized Bessel functions[\onlinecite{Reiss1980}] 
\bq
J_{n}(x,y)=\sum_{k=-\infty}^{\infty}J_{n-2k}(x)J_{k}(y).
\label{eq:genJ}
\eq
This brings (\ref{eq:psiVolkovperp}) to the Floquet form
\barr
\psi_{{\bf K},s}^{F}(\brho,z,t)
&=&
\exp\left[i\left({\bf K}\brho -\frac{(\hbar{\bf K})^2}{2m^{*}\hbar}t-\frac{E_{s}+U_{p}}{\hbar}t\right)\right]\nonumber\\
&\times&
\sum_{n=-\infty}^{\infty} e^{-i n\omega_{s}t}(i)^{n}J_{n}\left(i Z_{s}\frac{\partial}{\partial z},\beta_{s}\right) u_{s}(z).
\label{eq:fullVolkov}
\earr

 Expression in the second line of (\ref{eq:fullVolkov}) has the appearance of a Fourier transform of a function (here also an operator in the $z$-space) periodic in the time interval $T=\omega_{s}/2\pi$.  Such components of the wave function may lead to periodic structures in the electron excitation spectra. The intensities of these structures are determined by the quantities $U_{p}$ and $Z_{s}$ which derive from ${\cal P}_{\perp}$ which, in turn, derives from the coupling (\ref{eq:Asperp}) of the electron with plasmons previously excited into a coherent state. Since the latter is expressed through the eigenvalues $\alpha_{Q}$ which depend on the history of plasmon pumping by external fields, the values of ${\cal P}_{\perp}$, and {\it a fortiori} of $U_{p}$ and $Z_{s}$, depend on external parameters of the present model.     

\section{Electron emission from surface Floquet bands}
\label{sec:Floquet}

Depending on the boundary conditions specific to a particular problem, the Volkov Ansatz-derived electron states may participate in emission or scattering processes as either initial, intermediate or final states.[\onlinecite{Keldysh1965,Faisal,Reiss1980,Madsen,Yalunin,Park,Gedik}] The initial field-dressed band states are conventionally termed Floquet or Bloch-Floquet states whereas the final outgoing field-dressed electron states are designated Volkov states.[\onlinecite{Yalunin,Park,Gedik}] In the present problem of electron emission from surface localized bands we assume their strong renormalization by the equally surface localized vector field (\ref{eq:bcalA}) that leads to their representation in the form (\ref{eq:fullVolkov}). On the other hand, we assume the outgoing delocalized electron emission states negligibly affected by the surface localized SP field (\ref{eq:bcalA}). This puts the present problem in close correspondence with the case of multiple absorption of photons treated in Ref. [\onlinecite{Faisal}]. 
To reveal the correspondence we first draw the analogy between the herein defined wavefunctions $\phi_{{\bf K},s}$, Eq.(\ref{eq:phi_s}), and $\psi_{{\bf K},s}^{F}$, Eq. (\ref{eq:fullVolkov}), and the wavefunctions (8) and (9) from Ref. [\onlinecite{Faisal}] that serve as input for the amplitudes defined in Eq. (10) therein. Next we observe the analogy between $H_{1}(t)$ of Ref. [\onlinecite{Faisal}] and the present electron-plasmon field interaction ${\cal V}(t)$ derived from (\ref{eq:V_A}) by substituting $\bcalA_{\perp}(z_{s},t)$ in the place of ${\bf A(r)}$. This produces 
\bq
{\cal V}(t)=\frac{{\cal P}_{\perp}^{2}}{2m_{s}}\sin^{2}(\omega_{s}t) -\frac{{\cal P}_{\perp}}{m_{s}}
\sin(\omega_{s}t)\hat{p}_{z},
\label{eq:calV}
\eq 
with $\hat{p}_{z}$ given in (\ref{eq:p_z}). Using this we can follow Ref. [\onlinecite{Faisal}] and write for the amplitude of vertical plasmonically induced electron transition ${\bf K}_{f}={\bf K}_{s}={\bf K}$ to a final outgoing wave $|\phi_{{\bf K},f}\rangle$
\bq
T_{{\bf K},f\leftarrow s}= -\frac{i}{\hbar}\int_{-\infty}^{\infty}dt\langle\phi_{{\bf K},f}(t)|{\cal V}(t)|\psi_{{\bf K},s}^{F}(t)\rangle,  
\label{eq:Tfi} 
\eq
where we take the effective electron mass in the outgoing state $|\phi_{{\bf K},f}\rangle$ equal to the bare mass $m$. Observe that (\ref{eq:Tfi}) describes the situation which is in contrast to the case of strong final state electron coupling to spatially homogeneous EM fields. It is also in contradistinction to the situation of multiquantum-induced electronic transitions between two Q2D surface bands (e.g. the SS and IP bands) which undergo Autler-Townes splitting by the external EM field.[\onlinecite{MarcelNComm}]  
On noticing that for vertical transitions the temporal dependence of the integrand in (\ref{eq:Tfi}) is of the form
\bq
e^{i\left(\frac{\hbar^{2}K^2}{2}\left(\frac{m^{*}-m}{m m^{*}}\right)+E_{f}-E_{s}\right)t/\hbar}\left(-\frac{i}{\hbar}\right){\cal V}(t)\exp\left[-\frac{i}{\hbar}\int^{t}{\cal V}(t')d t'\right],
\label{eq:integrand}
\eq
we first integrate the RHS of (\ref{eq:Tfi}) by parts assuming the scattering boundary conditions ${\cal V}(t\rightarrow \pm\infty)\rightarrow 0$ and use (\ref{eq:fullVolkov}) to obtain the dimensionless transition amplitude
\barr
T_{{\bf K},f\leftarrow s}&=& -2\pi i\left(\frac{\hbar^{2}K^2}{2}\left(\frac{m^{*}-m}{m m^{*}}\right)+E_{f}-E_{s}\right)\nonumber\\
&\times&
\sum_{n=-\infty}^{\infty} i^{n}\langle\phi_{{\bf K},f}|J_{n}\left(i Z_{s}\frac{\partial}{\partial z},\beta_{s}\right)|\phi_{{\bf K},s}\rangle\nonumber\\
&\times&
\delta\left(\frac{\hbar^{2}K^2}{2}\left(\frac{m^{*}-m}{m m^{*}}\right)+E_{f}-E_{s}-U_{p}-n\hbar\omega_{s}\right)
\label{eq:Tfin}
\earr
in which the phases between successive transitions $n\rightarrow n\pm 1$ change by $\pm\pi/2$.
  
The large energy differences between different $n$-contributions to (\ref{eq:Tfin}) prevent their constructive or destructive interference in the total transition probability $|T_{{\bf K},f\leftarrow s}|^{2}$. Expressing one of the $\delta$-functions in this square through the equivalent Kronecker symbol representation 
\bq
\delta(E_{k}-E_{k'})\rightarrow \frac{L_{z}}{2\pi}\delta_{k,k'}\left/\left(\frac{\partial E_{k}}{\partial k}\right)\right.
\label{eq:Kronecker}
\eq
where $L_{z}$ denotes the quantization length in $z$-direction (cf. Eq. (261) in Ref. [\onlinecite{PhysRep}]), we may write 
\bq
|T_{{\bf K},f\leftarrow s}|^{2}=\sum_{n=-\infty}^{\infty}|T_{{\bf K},k_{f}\leftarrow s}^{(n)}|^{2}=\sum_{n=-\infty}^{\infty}W_{{\bf K},f\leftarrow s}^{(n)}2\pi\rho(E_{k}^{(n)})
\label{eq:Wtot}
\eq
where   
\barr
 W_{{\bf K},f\leftarrow s}^{(n)}&=& \frac{2\pi}{\hbar} \left(\frac{\hbar^{2}K^{2}}{2}\left(\frac{m^{*}-m}{m m^{*}}\right)+E_{k_{f}}-E_{s}\right)^{2}\nonumber\\
&\times&
\left|\langle\phi_{{\bf K},f}|J_{n}\left(i Z_{s}\frac{\partial}{\partial z},\beta{s}\right)|\phi_{{\bf K},s}\rangle\right|^{2}\nonumber\\
&\times&
\delta\left(\frac{\hbar^{2}K^{2}}{2}\left(\frac{m^{*}-m}{m m^{*}}\right)+E_{k_{f}}-E_{s}-U_{p}-n\hbar\omega_{s}\right).
\label{eq:wfin}
\earr
plays the role of transition rate, and 
\bq
\rho(E_{k_{f}}^{(n)})=\frac{L_{z}}{2\pi}\left/\left(\frac{\partial E_{k_{f}}^{(n)}}{\partial k_{f}}\right)\right.
\label{eq:rho}
\eq
 is the density of $k$-states around the final state energy $E_{k_{f}}^{(n)}$ in the $n$-th side band. Expressions (\ref{eq:wfin}) and (\ref{eq:rho}) are of reciprocal dimension which renders (\ref{eq:Wtot}) dimensionless.

%-----------------  Fig. 2 -----------------------
\begin{figure}[tb]
\rotatebox{0}{\epsfxsize=8 cm \epsffile{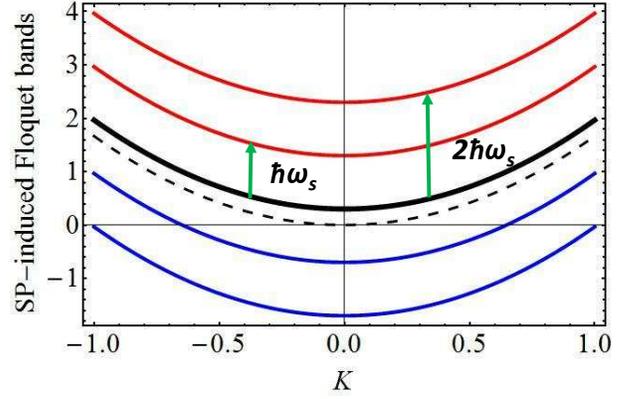}} 
\caption{Schematic of the SP-induced 2D Floquet band structure (arbitrary units) described in the effective mass approximation, as predicted by expression (\ref{eq:fullVolkov}) and manifesting in Eq. (\ref{eq:wfin}). Thin dashed black curve denotes the unperturbed 2D band and full thick black curve its ponderomotive shifted replica ($n=0$). Red and blue curves denote plasmonic Floquet side bands for positive and negative $n$, respectively. Vertical arrows exemplify one and two SP-assisted electronic transitions induced by the action of potential ${\cal V}(t)$ on the Volkov-dressed electronic state $|\psi_{{\bf K},s}^{F}(t)\rangle$ in Eq. (\ref{eq:Tfi}).  }
\label{SPFloquet}
\end{figure}
%------------------------------------------------------

The structures of closed form solutions for the transition amplitude (\ref{eq:Tfin}) and rate (\ref{eq:wfin}) clearly reveal the plasmon field driven surface electronic Floquet bands shifted from the parent one by the positive ponderomotive energy $U_{p}$ and multiples of $\hbar\omega_{s}$. According to (\ref{eq:wfin}) and (\ref{eq:genJ}) the weight of each $n$-th sideband is determined by an infinite sequence of combinations of one- and two-plasmon assisted processes generated by the linear and quadratic electron-plasmon coupling from the interaction Hamiltonian (\ref{eq:V_A}).    
Figure \ref{SPFloquet} illustrates the schematic of energetics of Q2D surface Floquet bands plasmonically generated from a parent SS-band and obeying the energy conservations in electronic transitions described by the amplitude (\ref{eq:Tfin}) and rate (\ref{eq:wfin}).  
Hence, the primary pumping of plasmonic coherent states may give rise to non-Einsteinian emission signal at multiples of $\hbar\omega_{s}$ provided the final state $|\phi_{{\bf K},f}\rangle$ is an "inverse LEED" outgoing wave solution[\onlinecite{Pendry,Krasovskii}] for the potential (\ref{eq:Vscal}). To illustrate this we first effectuate the operator $\partial/\partial z$ from the second argument of the generalized Bessel functions in action on the Fourier transform (FT) of the $z$-component of unperturbed initial SS-wavefunction
\bq
u_{s}(z)=\frac{1}{2\pi}\int dk_{z}e^{ik_{z}z}\tilde{u}_{s}(k_{z}). 
\label{eq:u(ks)}
\eq
 Next, we approximate the final state wavefunction corresponding to the energy $E_{k_{f}}=\hbar^{2}k_{f}^{2}/2m$ by the $L_{z}$-box normalized outgoing wave of unit amplitude[\onlinecite{Ashcroft}] 
\bq
u_{f}(z)\rightarrow e^{ik_{f}z}/\sqrt{L_{z}},
\label{eq:uf}
\eq
 that is unaffected by the strongly surface localized vector potential (\ref{eq:Az_s}). Using this normalization expression (\ref{eq:rho}) yields  $2\pi\rho(E_{k_{f}}^{(n)})=1/\hbar j_{z}^{(n)}$ where $j_{z}^{(n)}=\hbar k_{f}^{(n)}/mL_{z}$ is the electron current in the $n$-th channel. Therefore, expression (\ref{eq:wfin}) describes electron emission current from the $n$-th channel in the same normalization. This is analogous to the photoemission current elaborated in Refs. [\onlinecite{Ashcroft,Adawi,Mahan}]. 
Then, for vertical transitions with $K=0$ the prefactors of $\delta$-functions on the RHS of (\ref{eq:wfin}) read
\bq 
{\cal W}_{f\leftarrow s}(k_{f},n)=
\frac{2\pi}{\hbar L_{z}}(E_{k_{f}}^{(n)}-E_{s})^2|\tilde{u}_{s}(k_{f})|^{2}w_{n}(-k_{f}Z_{s},\beta_{s}).
\label{eq:Tz}
\eq
where $w_{n}(-k_{f}Z_{s},\beta_{s})=|J_{n}\left(-k_{f}Z_{s},\beta_{s}\right)|^{2}$. Here the Fourier transform $\tilde{u}_{s}(k_{f})$ plays the role of static form factor for inelastic transitions whereas only $w_{n}(-k_{f}Z_{s},\beta_{s})$ depends on the dynamics of plasmon field through $Z_{s}$ and $\beta_{s}$ defined in (\ref{eq:Z_s}) and (\ref{eq:beta_s}), respectively.[\onlinecite{Jarguments}] The linear coupling result, in which the quadratic coupling ${\bf A^{2}(r)}/2m_{s}$ is neglected, is obtained from the term $n=1$ and $k=0$ in the expansion (\ref{eq:genJ}). In the first order perturbation with linear coupling (Born approximation) the expression $w_{n}(-k_{f}Z_{s},\beta_{s})$ in (\ref{eq:Tz}) is replaced by $|-k_{f}Z_{s}/2|^{2}$. 

The static form factor $\tilde{u}_{s}(k_{f})$ from (\ref{eq:Tz}) acts as either a muffler or an amplifier  for the  transition amplitudes (\ref{eq:Tfin}) and rates (\ref{eq:wfin}) with $k_{f}$ and $E_{k_{f}}$ selected by the $\delta$-function on the RHS of these expressions. Moreover, since the generalized Bessel functions  have zeros on the real axis the variation of their arguments in (\ref{eq:Tz}) can give rise to resonant and antiresonant behaviour in the transition rate (\ref{eq:wfin}). This modulating effect is caused by the interference among the various intermittent plasmon absorption and emission processes that lead to the same $n$-th final Floquet state.
If the preference of electron excitation from the Fermi level found for bulk systems[\onlinecite{plasPE,Hopfield1965}] holds also at surfaces, such yields would manifest as discernible peaks in the electron emission spectra.

Expressions for the transition amplitude (\ref{eq:Tfin}) and rate (\ref{eq:wfin}) provide a proof of concept for non-Einsteinian electron emission from surface Floquet bands generated by sufficiently populated plasmon clouds prepumped in in interactions of external EM fields with electrons in metals.[\onlinecite{plasPE}] They were obtained within the framework of dipole approximation $Q\rho\ll 1$ for the plasmonic field which allowed a rather straightforward navigation among the various intermediate steps of derivation. The forms of plasmonic vector potentials beyond the the dipole approximation are presented in  Appendix \ref{sec:beyondDA}.

The results (\ref{eq:Tfin}) and (\ref{eq:wfin}) are based on the interaction (\ref{eq:V_A}) and the issue remains as how much they obey gauge-invariance.[\onlinecite{Boyd2004}] Incidentally, assessments of the gauge invariance have been made for electron interactions with homogeneous EM fields in solids and the results showed equivalence of the two gauges.[\onlinecite{SchuelerPRB,SchuelerJESRP}]

\section{Results and discussion}
\label{sec:discussion}

Signatures of Floquet sidebands predicted by the RHS of expression (\ref{eq:wfin}) largely depend on two factors. The first is related to the values of  generalized Bessel functions $J_{n}(x,y)$ with parametric variables $\beta_{s}$ and $Z_{s}$ characterizing the interacting electron-SP system. The second one pertains to the overall magnitude of (\ref{eq:Tz}) which determines the weight of each $\delta$-function in the sum over the Floquet band index $n$ in (\ref{eq:wfin}). Both quantities depend on ${\cal P}_{\perp}$ which can be readily calculated from (\ref{eq:calP}) once ${\cal A}_{Q}$ defined in Eq. (\ref{eq:AQ}) and $z_s$ are known. To facilitate the  ${\bf Q}$-summation over modes in (\ref{eq:calP}) we introduce an effective plasmonic coherent state amplitude $\alpha$ through the Ansatz[\onlinecite{Eliashberg,Grimvall}]
\bq
\sum_{\bf Q}\alpha_{Q}{\cal A}_{Q}e^{-Q|z_{s}|}\rightarrow \alpha \sum_{\bf Q}{\cal A}_{Q}e^{-Q|z_{s}|}.
\label{eq:alphaAnsatz}
\eq
Tunable $\alpha$ serves as a measure of the efficiency of pumping the SP coherent state[\onlinecite{plasPE}] and its factorization enables a straightforward evaluation of (\ref{eq:calP}). This  gives in atomic units of length $a_{B}$, energy $e^{2}/a_{B}=1\mbox{H}$, momentum $\hbar/a_{B}$ the $\alpha$-scaled parameters[\onlinecite{scaling}] determining (\ref{eq:Wtot})  
\barr
{\cal P}_{\perp}&=&\frac{3}{8} \alpha 
\left(\frac{\hbar\omega_{s}}{1\mbox{H}}\right)^{-1/2}\left(\frac{z_{s}}{a_{B}}\right)^{-5/2}\left(\frac{\hbar}{a_{B}}\right),
\label{eq:Pperp}\\
U_{p}&=&\frac{9}{256}\alpha^{2}\left(\frac{m}{m_{s}}\right)\left(\frac{\hbar\omega_{s}}{1 \mbox{H}}\right)^{-1} \left(\frac{z_{s}}{a_{B}}\right)^{-5}\times 1\mbox{H},     
\label{eq:Ealpha}\\ 
\beta_{s}&=&\frac{9}{512}\alpha^{2}\left(\frac{m}{m_{s}}\right)\left(\frac{\hbar\omega_{s}}{1 \mbox{H}}\right)^{-2} \left(\frac{z_{s}}{a_{B}}\right)^{-5},
\label{eq:beta}\\      
Z_{s}&=&\frac{3}{8}\alpha \left(\frac{m}{m_{s}}\right)\left(\frac{\hbar\omega_{s}}{1 \mbox{H}}\right)^{-3/2} \left(\frac{z_{s}}{a_{B}}\right)^{-5/2}a_{B}.
\label{eq:Zalpha}
\earr  

%-----------------  Fig. 3 -----------------------
\begin{figure}[tb]
\rotatebox{0}{\epsfxsize=8.3 cm \epsffile{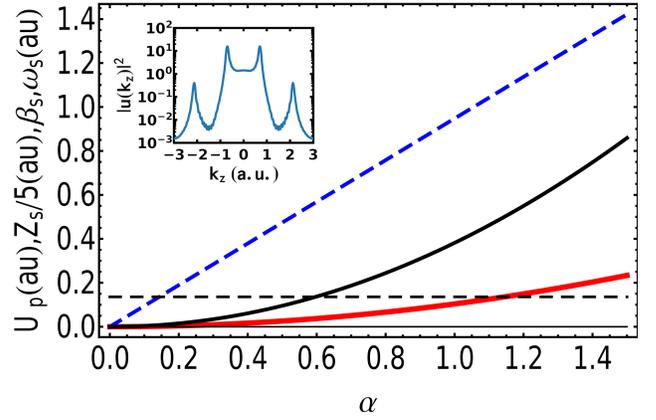}} 
\caption{$\alpha$-scaling of entries in expression (\ref{eq:Tz}) that determines the transition rate (\ref{eq:wfin}):  $U_{p}$ (lower red line), $\beta_{s}$ (upper black line) and rescaled $Z_{s}/5$ (upper blue dashed line) as functions of the plasmonic coherent state parameter $\alpha$ and other parameters fixed at the values characteristic of Ag(111) surface (see text). Horizontal dashed black line denotes the value of SP energy $\hbar\omega_{s}$. Vertical scale in atomic units. Inset: Static form factor $|u_{z}(k_{f})|^{2}$, Eq. (\ref{eq:u(ks)}), as function of the final state wavevector $k_{f}$.}
\label{ponderomotives}
\end{figure}
%------------------------------------------------------

We are now in a position to make semiquantitative estimates of (\ref{eq:Pperp}) and the ensuing quantities (\ref{eq:Ealpha}), (\ref{eq:beta}) and (\ref{eq:Zalpha}) for the parent SS-state on Ag(111) surface. We compute $u_{s}(z)$  using the DFT methods described in Refs. [\onlinecite{ACSPhotonics,plasPE,AndiPRB22}] and obtain $E_{s}=-0.081\mbox{eV}=-0.003$ H relative to the Fermi level. With the previously estimated $\hbar\omega_{s}\simeq 3.7 \mbox{eV}\simeq 0.136$ H and $z_{s}\simeq 1.1-1.2$ $a_{B}$[\onlinecite{Liebsch}], and the effective SS-state electron mass for perpendicular motion $m_{s}\simeq m$ we can plot the Floquet parameters (\ref{eq:Ealpha}), (\ref{eq:beta}) and (\ref{eq:Zalpha}) as functions of the effective plasmonic coherent state parameter $\alpha$. The results for Ag(111) surface are presented in Fig. \ref{ponderomotives}. This determines the $\alpha$-dependence of the main constituent of expression (\ref{eq:Tz}), viz.  
\bq
w(n,\alpha)=w_{n}(-k_{f}^{(n)}Z_{s},\beta_{s}), 
\label{eq:wnalpha}
\eq
where $k_{f}^{(n)}$ is constrained to the energy shell of $n$-plasmon absorption processes expressed through the $\delta$-functions on the RHS of (\ref{eq:wfin}). Thus, for two-plasmon assisted electron emission from clean Ag(111) surface $E_{f}^{(2)}$ and $k_{f}^{(2)}$ should correspond to the situation depicted in Fig. \ref{AllAg}. For one-plasmon case $E_{f}^{(1)}$ and $k_{f}^{(1)}$ should correspond to emission from Ag(111) surface with reduced work function, e.g. by alkali submonolayer adsorption.[\onlinecite{Horn,Petek2008,PetekJPC2011,Raseev}]

\begin{figure}[tb]
\rotatebox{0}{\epsfxsize=8.5 cm \epsffile{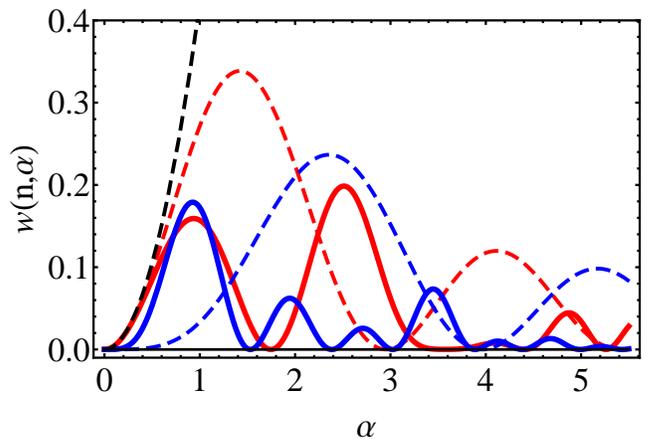}} 
\caption{ $\alpha$-dependence of $w(n,\alpha)$ defined in Eq. (\ref{eq:wnalpha}) that determine the transition rates (\ref{eq:wfin}) of plasmonically assisted electron emissions from SS-derived Floquet bands on Ag(111). 
Thick red curve: $w(1,\alpha)$ for full linear and quadratic electron coupling to SP vector potential $\bcalA$. Dashed red curve: $w(1,\alpha)$ for linear coupling. Dashed black curve: $w(1,\alpha)$ in the first order Born approximation with linear coupling. Thick blue curve: $w(2,\alpha)$ for full linear and quadratic  coupling. Dashed blue curve: $w(2,\alpha)$ for linear coupling. }
\label{wnalpha}
\end{figure}

The behaviour of $w(n,\alpha)$ for $n=1$ and $n=2$ SP-assisted electron emission from Q2D Floquet states calculated for the full linear and quadratic coupling (i.e. coupling to both $\bcalA_{\perp}$ and $\bcalA^{2}_{\perp}$), only linear coupling to $\bcalA_{\perp}$, and in the first order Born approximation, is illustrated in Fig. \ref{wnalpha} as function of $\alpha$. Here for the sake of comparing our model results we relate $k_{f}^{(n)}$ to Ag(111) surface with work function reduced by $\Delta\phi=1.3$ eV [\onlinecite{Horn}]. In doing so we assert that the initial SS-states remain robust with respect to $\Delta\phi$ (cf. Figs. 1 in [\onlinecite{Petek2008,PetekJPC2011,Raseev}]) whereas the IP-states downshift from the energy interval of plasmonically driven $|SS\rangle\rightarrow|f\rangle$ transition resonances. 

The results presented in Fig. \ref{wnalpha} convey several important messages: 
\\
{\it (i)}  $w(n,\alpha)$ are extremely sensitive to the effective number of plasmons (as measured by $\alpha$) that have been pumped into the exciting coherent state. {\it 
\\
(ii)} For the studied component of SP vector potential the interplay between the linear and quadratic coupling is very strong so that they must be treated on an equivalent footing.[\onlinecite{eqfooting,Despoja2016}] Figure \ref{wnalpha} specifically illustrates this for one- and two-plasmon-driven electron emissions. 
\\
{\it (iii)} Quadratic coupling strongly renormalizes also one-plasmon induced transitions (cf. the difference between the full and thin dashed red curves), in a fashion analogous to the Debye-Waller (DW) factor.[\onlinecite{nonlinscat}] The first order Born approximation result deviates from the linear coupling for $\alpha>1/2$.   
\\
{\it (iv)} Quadratic coupling gives an overwhelmingly dominant contribution to instantaneous two-plasmon assisted electron emissions for $\alpha<1$ (cf. full thick and thin dashed blue lines). The successive one-plasmon assisted processes (full red curve) start to dominate over the  two plasmon ones (thin dashed blue line) for $\alpha>1$. 
\\
{\it (v)} Fully renormalized one- and two-plasmon driven electron emissions exhibit a resonant-like behaviour for the value $\alpha\sim 1/2$ which according to (\ref{eq:alphaAnsatz}) corresponds to an evenly distributed subsingle occupation of  modes in the primary pumped plasmonic coherent state. 

 The pumping of coherent plasmonic states with optimal properties[\onlinecite{plasPE}] represents one of the limiting factors for observation of electron emission from the plasmonic Floquet bands. Another limitation affecting expression (\ref{eq:Tz}), and thereby (\ref{eq:wfin}), may come from the form factor for higher order processes that result in large $k_{f}^{(n)}$ and correspondingly small $|\tilde{u}_{s}(k_{f}^{(n)})|^{2}$ (cf. inset in Fig. \ref{ponderomotives}).

%-----------------  Fig. 4 -----------------------
%------------------------------------------------------

\begin{figure}[tb]
\rotatebox{0}{\epsfxsize=8.5 cm \epsffile{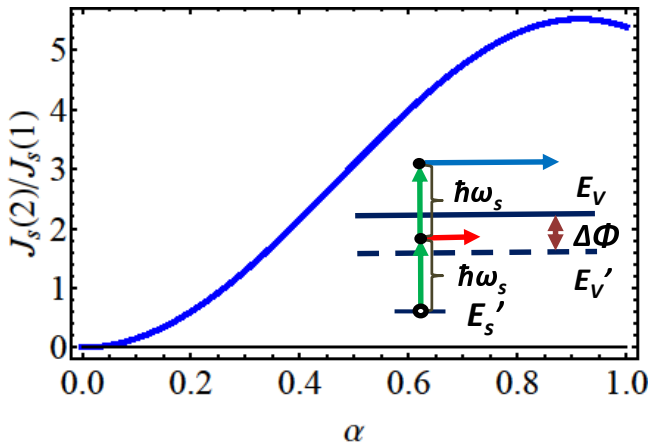}} 
\caption{ Relative intensities of two- and one-plasmon driven electron emissions ${\cal J}_{s}(2)/{\cal J}_{s}(1)$, defined in (\ref{eq:P21}), from the same initial SS-state on Ag(111) surface with energy $E_{s}'=E_{s}+U_{p}$ and ${\bf K}=0$ plotted as a function of the plasmonic coherent state parameter $\alpha$.  Inset: Blue and red horizonatal arrows illustrate two- and one-plasmon driven electron emissions from  $E_{s}'$-level to states above the vacuum level $E_{V}'$ of the Ag(111) surface with work function reduced by $\Delta\phi\approx 1.3$ eV[\onlinecite{Horn}] (dashed black line). Scaling of arrow lengths corresponds to $k_{f}^{(2)}/k_{f}^{(1)}=2.14$. This provides complementary information required for full characterization of the SP field-driven electron emission. }
\label{P21}
\end{figure}

Adaptation of expressions (\ref{eq:Tfin}) for description of experimental situation of $n$-plasmon assisted electron emission with ${\bf K}=0$ proceeds by combining expressions (\ref{eq:Kronecker}), (\ref{eq:wfin}), (\ref{eq:rho}) and (\ref{eq:Tz}) to obtain the electron emission current ${\cal J}_{s}(k_{f}^{(n)},n)$ from the narrow energy interval around $E_{f}^{(n)}$ in the $n$-th Floquet sideband of the parent $s$-band. This gives  
\bq
{\cal J}_{s}(k_{f}^{(n)},n)={\cal W}_{f\leftarrow s}(k_{f}^{(n)},n)\rho(E_{f}^{(n)}).
\label{eq:Pn}
\eq
This expression is independent of the quantization length $L_{z}$ and has the dimension of velocity (i.e. of the current).
Using this the assessment of $\alpha$ can be attempted in the measurements of one- and two-plasmon driven electron emissions from one and the same surface with sufficiently reduced workfunction ($\phi_{\rm red}<\hbar\omega_{s}$) (see inset in Fig. \ref{P21}). Here the intensities of one- and two-plasmon induced peaks may enable the estimate of $\alpha$ through the comparison of relative values of experimental emission intensities with the corresponding theoretical predictions for $f\leftarrow s$ transition probabilities defined by  
\barr
\frac{{\cal J}_{s}(2)}{{\cal J}_{s}(1)}&=& {\cal J}_{s}(k_{f}^{(2)},2)/{\cal J}_{s}(k_{f}^{(1)},1)\nonumber\\
&=&
\frac{(E_{f}^{(2)}-E_{s})}{(E_{f}^{(1)}-E_{s})}\left(\frac{|\tilde{u}_{s}(k_{f}^{(2)})|^{2}/k_{f}^{(2)}}{|\tilde{u}_{s}(k_{f}^{(2)})|^{2}/k_{f}^{(1)}}\right)\frac{w(2,\alpha)}{w(1,\alpha)}
\label{eq:P21}
\earr
and illustrated in the main body of Fig. \ref{P21}.[\onlinecite{alpha/2}]
In the present situation where $w(2,\alpha)$ and $w(1,\alpha)$ are of nearly the same magnitude (cf. Fig. \ref{P21}) the enhancement of ${\cal J}_{s}(2)$ over ${\cal J}_{s}(1)$ for $\alpha>1/3$ arises from the channel kinetics ratio 
\bq
\eta_{f\leftarrow s}(2,1) =\frac{|\tilde{u}_{s}(k_{f}^{(2)})|^{2}/k_{f}^{(2)}}{|\tilde{u}_{s}(k_{f}^{(1)})|^{2}/k_{f}^{(1)}},
\label{eq:etaF}
\eq
that is retrivable from insets in Figs. \ref{ponderomotives} and \ref{P21}. 
With the above developed prerequisites the estimates of $\alpha$ may provide insightful and much desired information on the generation of SP coherent states upon irradiation of surfaces by strong trans-resonant EM fields.[\onlinecite{plasPE}]

%-----------------  Fig. 5 -----------------------
%------------------------------------------------------

\section{Summary and conclusions}
\label{sec:conclusion}     

Starting from a simple model of electron-SP interaction we have demonstrated  that the vector potential associated with the pre-generated coherent surface plasmon field can give rise to Q2D Floquet bands that build on the parent metallic surface state band.
Based on the nonperturbative Volkov Ansatz-like dressing of electron wavefunctions we have derived general expressions for the rates (\ref{eq:wfin}) and (\ref{eq:Tz}), and currents (\ref{eq:Pn}) describing SP-driven electron emission from surface Floquet bands.  The entries in these expressions as formulated in (\ref{eq:u(ks)}), (\ref{eq:Ealpha}), (\ref{eq:beta}) and (\ref{eq:Zalpha}) are evaluated and displayed in Figs. \ref{ponderomotives}, \ref{wnalpha} and \ref{P21} for the parent SS-band on Ag(111) surface. This enables their direct reading from the plots and substitution into (\ref{eq:wfin}), (\ref{eq:Tz}) and (\ref{eq:Pn}) for the final assessment of multiplasmon driven electron yields and their relative intensities (\ref{eq:P21}) exemplified in Fig. \ref{P21}. Their overall magnitudes turn out very sensitive to the characteristics of pre-generated plasmonic coherent state embodied in its effective amplitude $\alpha$ defined in (\ref{eq:alphaAnsatz}), as well as on the static form factor (\ref{eq:u(ks)}) arising from the parent surface band structure. 

The obtained results are consistent with the experimental facts pertaining to the observed non-Einsteinian photoemission from Ag surfaces.[\onlinecite{MarcelPRL,ACSPhotonics,plasPE,Horn}] Thereby they indicate the need for and the power of nonperturbative solutions for wavefunctions describing electron motion in strong plasmonic vector fields. This invites extensions of the earlier studies of electron dynamics subject to strong periodic perturbations[\onlinecite{FaisalKaminski97,FaisalKaminski05,Park,PengZhang}] to the surface geometry and vector fields discussed in the present work. Equally inviting is the experimental search for the systems with the discussed Floquet properties. Particularly revealing would be the  assessment of intensities of the one- and two-plasmon driven electron yields from one and the same surface. We propose that the relative intensities (\ref{eq:P21}) of such non-Einsteinian peaks be identified with the fingerprints of effective amplitudes $\alpha$ of the exciting plasmonic fields. This property opens the posibility of monitoring and controlling the excitation of plasmonic fields. To this end the results outlined in Sec. \ref{sec:discussion} may provide useful guidelines for making contact between experiment and theory because the relative intensities shown in Fig. \ref{P21} can be recalculated for other surfaces as well. It is also envisaged that the developed theory could be extended to the studies of plasmonically driven electron excitations in other geometries and in nanoparticles.[\onlinecite{Atwater14,Nordlander15,Linic,Louie15,Halas15,Boriskina,Narang2016,deAbajo16,Atwater16,Khurgin,Govorov17}]  

%APPENDIX  ================================================================

\appendix
\section{Plasmonic vector potential beyond the dipole approximation}
\label{sec:beyondDA} 
  
The plasmonic vector potential (\ref{eq:Az_s}) beyond the dipole approximation is calculated from expressions (\ref{eq:Aspar}) and (\ref{eq:Asperp}) by projecting them, respectively, onto the vectors $\hat{\bf e}_{\bf K}={\bf K}/K$ and $\hat{\bf e}_{z}$  in conjunction with which they appear in the Volkov Ansatz. Denoting by $\theta$ the angle between the vectors ${\bf K}$ and $\brho$ and by $\varphi$ between the vectors ${\bf Q}$ and $\brho$ we perform angular integration $\int_{0}^{2\pi}\dots d\varphi/2\pi$ to find
\barr
 \hat{\bf e}_{\bf K}\bcalA_{\parallel}(\brho,\bar{z}_{e},t)&=&\frac{L^2}{2\pi} \int_{0}^{\infty}Q dQ \alpha_{Q}{\cal A}_{Q}e^{-Q|z_{s}|}J_{1}(Q\rho)\nonumber\\
&\times&
\cos\theta\sin(\omega_{s}t),
\label{eq:KA}
\earr
where $J_{1}(Q\rho)$ is the Bessel function of the first kind and first order. Note also in passing that according to (\ref{eq:AQ}) we have ${\cal A}_{Q}\propto 1/L^{2}$ which cancels out the same factor in front of the inegral.

Using the same procedure we obtain
\bq
\hat{\bf e}_{z}\bcalA_{\perp}(\brho,\bar{z}_{e},t)=\frac{L^2}{2\pi} \int_{0}^{\infty}Q dQ \alpha_{Q}{\cal A}_{Q}e^{-Q|z_{s}|}J_{0}(Q\rho)\sin(\omega_{s}t)
\label{eq:ezA}
\eq
which in the dipole approximation leads to Eq. (\ref{eq:calP}).

Analogously we obtain the terms quadratic in the plasmon vector field  beyond the dipole approximation. Denoting by $\varphi_{1}$ and $\varphi_{2}$ the angles between the plasmon wavevectors ${\bf Q}_{1}$ and ${\bf Q}_{2}$, respectively,  and $\brho$, we find  
\bq
\bcalA_{\parallel}^{2}(\brho,z_{s},t)=\left(\frac{L^{2}}{2\pi}\int_{0}^{\infty}Q dQ\alpha_{Q}{\cal A}_{Q}e^{-Q|z_{s}|}J_{1}(Q\rho)\right)^{2}\sin^{2}(\omega_{s}t).
\label{eq:Apar^2}
\eq
 Likewise we obtain from (\ref{eq:ezA}) the expression 
\bq
\bcalA_{\perp}^{2}(\brho,z_{s},t)=\left(\frac{L^2}{2\pi} \int_{0}^{\infty}Q dQ \alpha_{Q}{\cal A}_{Q}e^{-Q|z_{s}|}J_{0}(Q\rho)\right)^{2}\sin^{2}(\omega_{s}t)
\label{eq:AAperp}.
\eq

The maximum value of $Q$ in the above integrals is effectively limited either by $1/z_{s}$ or the maximum plasmon wavevector $Q_{c}$, whichever is smaller. The Bessel functions in the integrands on the RHS of (\ref{eq:KA}), (\ref{eq:ezA}), (\ref{eq:Apar^2}) and (\ref{eq:AAperp}) behave for small $Q\rho$ as 
\barr
J_{0}(Q\rho)&=& 1-{\cal O}\left((Q\rho)^{2}/4\right),\nonumber\\
J_{1}(Q\rho)&=& {\cal O}\left(Q\rho/2\right).
\label{eq:Bessel}
\earr
Therefore in the dipole approximation $Q\rho\ll 1$ only expressions (\ref{eq:ezA}) and (\ref{eq:AAperp}) produce $\rho$-independent contributions which dominate all others. This justifies their use in construction of the Volkov wavefunction (\ref{eq:fullVolkov}) and its subsequent representation through ${\cal P}_{\perp}$ and ${\cal E}_{\perp}$, ultimately leading to the Floquet band appearance of expressions (\ref{eq:Tfin}) and (\ref{eq:wfin}).

%%%%%%%%%%%%%%%%%%%%%%%%%%%%%%%%%%%%%%%%%%%%%%%%%%%%%%%%%%%%%%%%%%%%


\begin{thebibliography}{11}
\bibitem{Truscott1} The validity of Volkov Ansatz for a general class of surface potentials was later demonstrated in Ref. [\onlinecite{Truscott2}]. For the special case of Coulomb potential it was demonstrated in Ref. [\onlinecite{Faisal}], see Eqs. (4) and (6) therein.
\bibitem{Glauber} These states are completely analogous to photonic coherent states excited by electronic motion (current) as described by Eqs. (9.16)-(9.23) in Ref. [\onlinecite{GlauberPR}].
\bibitem{scaling} The displayed $\alpha$-scaling of expressions (\ref{eq:Pperp})-(\ref{eq:Zalpha}) arises from and signifies the excitation dynamics of the coupled electron-SP system. This is in contrast to the system ground state energy and related quantities which scale with the interelectron spacing parameter $r_{s}$.
\bibitem{alpha/2} This may be contrasted with ${\cal J}^{(Poiss)}(2,1)\propto\alpha^2/2$ obtained from (\ref{eq:wfin}) when it follows Poissonian distribution. Such distributions are generated in the linear coupling regime by cumulant expansion treatment of consecutive one-plasmon absorptions without any intermittent re-emissions or vice versa.[\onlinecite{GW+C}]
\end{thebibliography}
\end{document}